\documentclass[11pt]{article}
\usepackage{graphicx}
\usepackage{amssymb,amsmath}
\usepackage[sort&compress,numbers]{natbib}
\usepackage{xcolor}

\begin{document}

\title{
\textbf{Corrections to scaling in the 3D Ising model:\\
a comparison between MC and MCRG results}
}

\author{J. Kaupu\v{z}s$^{1,2,3}$ 
\thanks{E--mail: \texttt{kaupuzs@latnet.lv}} \hspace{1ex}, 
R. V. N. Melnik$^{3,4}$ \\
$^1$ Laboratory of Semiconductor Physics,
Institute of Technical Physics, \\ Faculty of Materials Science and Applied Chemistry,
Riga Technical University, \\ P. Valdena 3/7, Riga, LV-1048, Latvia,
kaupuzs@latnet.lv\\
$^2$ Institute of Science and Innovative Technologies, \\
University of Liepaja, 14 Liela Street, Liepaja LV--3401, Latvia \\
$^3$ The MS2 Discovery Interdisciplinary Research Institute, \\
Wilfrid Laurier University, Waterloo, Ontario, Canada, N2L 3C5, 
rmelnik@wlu.ca \\
$^4$ BCAM - Basque Center for Applied Mathematics, E48009 Bilbao, Spain}

\maketitle

\begin{abstract}
Corrections to scaling in the 3D Ising model are studied
based on Monte Carlo (MC) simulation results for very large lattices
with linear lattice sizes up to $L=3456$.
Our estimated values of the correction-to-scaling
exponent $\omega$ tend to decrease below the usually accepted
value about $0.83$ when the
smallest lattice sizes are discarded from the fits.
This behavior apparently confirms some of the known
estimates of the Monte Carlo renormalization group (MCRG)
method, i.~e., $\omega \approx 0.7$ and $\omega = 0.75(5)$.
We discuss the possibilities
that $\omega$ is either really smaller than usually expected
or these values of $\omega$ describe some transient behavior.
We propose refining MCRG simulations and analysis to resolve this
issue.
In distinction from $\omega$,
our actual MC estimations of the critical exponents
$\eta$ and $\nu$ provide stable values
$\eta=0.03632(13)$ and $\nu=0.63017(31)$, which
well agree with those of the conformal bootstrap
method, i.~e., $\eta=0.0362978(20)$ and $\nu=0.6299709(40)$.
\end{abstract}


\section{Introduction}
\label{intro}

The 3D Ising model is one of the most extensively studied three-dimensional systems.
Focusing on the non-perturbative methods, one has to mention numerous Monte Carlo (MC)
simulations -- see e.~g.~\cite{HasRev,Has1,Lundow,KMR17,Ferrenberg} and
references therein,
as well as Monte Carlo renormalization group (MCRG)
studies~\cite{Gupta,Ron,Chung} of the
3D Ising model.
High- and low-temperature series expansions~\cite{Wipf,HTCompostrini} are other numerical approaches,
dealing directly with the Ising model. The large mass expansion~\cite{Yamada}
can be mentioned as an alternative to the high temperature expansion.
There are many different analytical and numerical approaches,
including the perturbative~\cite{Amit,Ma,Justin,Kleinert,PV,GZJ,PS08,Shalaby} and the nonperturbative
renormalization group (RG) methods~\cite{Report1,Report2,WshPRE,ourNPRG,uuu},
as well as the conformal bootstrap method~\cite{Showk,bootstrap,CFTrecent},
which allow determining the critical exponents of the 3D Ising universality class.
The critical exponents of this universality class are assumed to be known
with a high precision, and, as usually claimed, the most accurate values are
provided by the conformal bootstrap method~\cite{Showk,bootstrap,CFTrecent}. A
reasonable consensus about
this has been reached, as regards the estimation of the critical exponents
$\eta$ and $\nu$.
The situation with the correction-to-scaling exponent $\omega$ is less clear. The common claim is that $\omega$
is about $0.83$, as consistent with the { results
$\omega=0.8303(18)$~\cite{Showk} and $\omega=0.82951(61)$~\cite{CFTrecent} of the conformal field theory (CFT) obtained via the conformal bootstrap method} and the MC estimate
$\omega=0.832(6)$ of~\cite{Has1}. However, the MCRG method has provided significantly
smaller $\omega$ values, i.~e., $\omega \approx 0.7$~\cite{Gupta} and
$\omega = 0.75(5)$~\cite{Ron}. Our earlier MC study~\cite{KMR17} has
also pointed to a possibly smaller $\omega$ value, however, with somewhat
large statistical errors.

Currently, we have extended and refined our MC simulations
for a better estimation of $\omega$, providing also an intriguing comparison
with the MCRG results. We discuss the related issues about
the correct value of $\omega$, as well as about refining MCRG simulations and analysis
for a more reliable estimation of $\omega$.

\section{The Monte Carlo simulation method and results}
\label{sec:MCrez}

We have simulated the 3D Ising model on simple cubic $L\times L \times L$ lattice with periodic boundary conditions and the Hamiltonian
\begin{equation}
H/T = - \beta \sum_{\langle ij \rangle} \sigma_i \sigma_j \;,
\end{equation}
where $T$ is the temperature measured in energy units, $\beta$ is the coupling constant
and $\langle ij \rangle$ denotes the pairs of
neighboring spins $\sigma_i = \pm 1$.
The MC simulations have been performed with the Wolff
single cluster algorithm~\cite{Wolff}, using its parallel implementation
described in~\cite{KMR_2010}. More precisely, we have used a hybrid parallelization,
where several independent simulation runs were performed in parallel, applying the parallel Wolff algorithm for each of them. Typically, four first simulation blocks (iterations described in~\cite{KMR_2010}) have been discarded to ensure
a very accurate equilibration. To save the simulation resources, only one run
has been performed at the beginning for large enough $L$, splitting the simulation
in several independent parallel runs only after the third iteration,
restarting the pseudo-random number generator for each new run.
Note that we have used two different pseudo-random number generators,
described and tested in~\cite{KMR_2010}, to verify the results.

The parallel Wolff algorithm of~\cite{KMR_2010} is based on the Open MP technique. It allows to speed up each individual run and increase the operational memory
available for this run. It is important for the largest lattice sizes from
the subset $L \le 3456$ we considered. We have also combined a slightly modified
parallel Wolff algorithm with spin coding in some cases to save the operational memory.
It means that a string of spin variables is encoded into a single
integer number. A modification of the parallel algorithm is necessary in this case to avoid the so-called critical racing conditions, where two or several processors try to write simultaneously
into the same memory init. The problem is solved by appropriately splitting the
wave-front of the growing Wolff cluster into parts
and using extra Open MP barriers
to ensure that only one of these parts is treated in parallel at a time.

We have essentially extended and refined our
earlier simulation results reported in~\cite{KMR17}, where the values of various
physical observables at certain pseudocritical couplings $\widetilde{\beta}_c(L)$
and $\bar{\beta}_c(L)$ have been obtained by the iterative method originally
described in~\cite{KMR_2010}. The pseudocritical coupling $\widetilde{\beta}_c(L)$
corresponds
to certain value $U=1.6$ (which is close to the critical value) of the ratio
$U=\langle m^4 \rangle / \langle m^2 \rangle^2$,
where $m$ is the magnetization per spin. The pseudocritical coupling $\bar{\beta}_c(L)$
corresponds to $\xi_{\mathrm{2nd}}/L=0.6431$, where 0.6431 is an approximate
critical value of this ratio~\cite{Has1}, $\xi_{\mathrm{2nd}}$ being the second moment correlation length (see~\cite{KMR17} for more details).
By this iterative method, the simulations are, in fact, performed at a set of
$\beta$ values in vicinity of the corresponding pseudocritical coupling.
It allows to evaluate the physical observables (several mean values)
at any $\beta$ near this pseudocritical coupling, using the Taylor series expansion.
In such a way, the final estimates of the quantities of interest at
$\widetilde{\beta}_c(L)$ or $\bar{\beta}_c(L)$
are obtained, including the value of the pseudocritical coupling itself.
There is some similarity of this method with the known tempering algorithms~\cite{Janke},
since the simulations are performed at slightly
different fluctuating $\beta$ values or inverse temperatures, which eventually
can lead to more reliable results.

In the current study, we have extended our earlier simulation results~\cite{KMR17} to $L=3456$ and $L=3072$ for the pseudocritical couplings
$\widetilde{\beta}_c(L)$ and $\bar{\beta}_c(L)$, respectively. The values of
$\widetilde{\beta}_c(L)$ and $\bar{\beta}_c(L)$ appear to be close to each other.
Moreover, the value of $U$ at $\beta=\bar{\beta}_c(L)$ is close to $1.6$,
e.~g., $U = 1.57252(16)$ at $L=16$, $U = 1.59380(32)$ at $L=64$ and even closer to $1.6$ at $L>64$. It allowed us to evaluate the physical observables at
$\beta=\widetilde{\beta}_c(L)$ for $16 \le L \le 3072$  from the simulations
with $\beta$ near $\bar{\beta}_c(L)$ by the Taylor series expansion
with negligibly small (as compared to one standard error) truncation errors. We have averaged over these results and those ones obtained
by directly using the pseudocritical coupling $\widetilde{\beta}_c(L)$ to obtain the refined final estimates
for $\beta=\widetilde{\beta}_c(L)$. We have used the weight factors
$\propto 1/\sigma_i^2$
in this averaging procedure, where $\sigma_i$ with $i=1,2$ are the
corresponding standard errors in these two cases. It ensures the minimal
resulting error.
Because of relatively long simulations with $\bar{\beta}_c(L)$,
this averaging procedure allowed us to improve significantly the
statistical accuracy of the final results for $\beta=\widetilde{\beta}_c(L)$.
The summary of our results at $\beta=\bar{\beta}_c(L)$ and
$\beta=\widetilde{\beta}_c(L)$ is provided in Tabs.~\ref{tab1} and~\ref{tab2},
respectively. In these tables, all those quantities are given, which are used in our following MC analysis.

\begin{table}
\caption{The pseudocritical couplings
$\bar{\beta}_c$ and the values of $U$,
$U_6=\langle m^6 \rangle / \langle m^2 \rangle^3$, $\chi/L^2$  (where
$\chi$ is the susceptibility) and $\partial Q/\partial \beta$
(where $Q=1/U$) at $\beta=\bar{\beta}_c$
depending on the linear system size $L$.
\label{tab1} }
\begin{center}
\begin{tabular}{|c|c|c|c|c|c|}
\hline
\rule[-2mm]{0mm}{7mm}
L & $\bar{\beta}_c$ & $U$ & $U_6$ &  $\chi/L^2$ & $10^{-3}\partial Q/\partial \beta$ \\
\hline
3072 & 0.2216546192(34) & 1.60437(179)& 3.1087(81) & 1.15780(98) & 285.54(2.51) \\
2560 & 0.2216546175(24) & 1.60334(89) & 3.1029(41) & 1.16405(68) & 216.33(76) \\
2048 & 0.2216546255(31) & 1.60222(96) & 3.0993(45) & 1.17543(54) & 151.62(64) \\
1728 & 0.2216546253(39) & 1.60410(84) & 3.1075(37) & 1.18184(47) & 116.08(42) \\
1536 & 0.2216546251(47) & 1.60270(93) & 3.1009(41) & 1.18663(46) & 96.42(44) \\
1280 & 0.2216546194(51) & 1.60165(71) & 3.0966(31) & 1.19490(40) & 71.62(21) \\
1024 & 0.2216546177(63) & 1.60223(54) & 3.0989(23) & 1.20460(33) & 50.36(13) \\
864  & 0.2216546253(81) & 1.60179(55) & 3.0980(24) & 1.21217(29) & 38.457(93)\\
768  & 0.2216546096(87) & 1.60272(52) & 3.1013(23) & 1.21672(30) & 32.064(72) \\
640  & 0.221654632(11)  & 1.60312(57) & 3.1023(25) & 1.22492(31) & 24.148(56) \\
512  & 0.221654616(15)  & 1.59984(56) & 3.0890(25) & 1.23525(33) & 16.705(41) \\
432  & 0.221654621(20)  & 1.60067(56) & 3.0922(25) & 1.24244(35) & 12.837(28) \\
384  & 0.221654607(22)  & 1.60058(61) & 3.0918(27) & 1.24796(32) & 10.627(25) \\
320  & 0.221654699(30)  & 1.60216(49) & 3.0980(21) & 1.25531(30) & 8.013(15) \\
256  & 0.221654685(44)  & 1.60004(50) & 3.0891(22) & 1.26533(30) &  5.578(12) \\
216  & 0.221654598(58)  & 1.59868(52) & 3.0828(23) & 1.27374(33) & 4.2620(82) \\
192  & 0.221654746(74)  & 1.59922(52) & 3.0850(23) & 1.27889(35) & 3.5512(72) \\
160  & 0.221654927(97)  & 1.59885(51) & 3.0835(22) & 1.28686(33) & 2.6626(49) \\
128  & 0.22165469(13)   & 1.59772(57) & 3.0780(25) & 1.29596(34) & 1.8657(37) \\
108  & 0.22165476(15)   & 1.59727(45) & 3.0778(27) & 1.30441(33) & 1.4246(25) \\
96   & 0.22165490(18)   & 1.59557(44) & 3.0681(19) & 1.30956(30) & 1.1810(17) \\
80   & 0.22165539(22)   & 1.59550(36) & 3.0672(16) & 1.31743(28) & 0.8872(12) \\
64   & 0.22165530(26)   & 1.59380(32) & 3.0593(14) & 1.32699(24) & 0.62241(66) \\
54   & 0.22165642(25)   & 1.59211(27) & 3.0513(12) & 1.33509(22) & 0.47606(53) \\
48   & 0.22165642(34)   & 1.59104(25) & 3.0462(11) & 1.33972(21) & 0.39478(39) \\
40   & 0.22165743(41)   & 1.58936(26) & 3.0381(11) & 1.34712(18) & 0.29639(28) \\
32   & 0.22166131(53)   & 1.58643(23) & 3.02425(94) & 1.35626(18) & 0.20828(17) \\
27   & 0.22166437(60)   & 1.58377(23) & 3.01220(93) & 1.36289(16) & 0.15901(13) \\
24   & 0.22166676(66)   & 1.58148(20) & 3.00122(83) & 1.36696(15) & 0.131954(91)\\
20   & 0.22167152(83)   & 1.57787(20) & 2.98419(84) & 1.37295(15) & 0.098941(62) \\
16   & 0.2216823(12)    & 1.57252(16) & 2.95893(65) & 1.37981(13) & 0.069608(33) \\
12   & 0.2217109(14)    & 1.56377(13) & 2.91737(54) & 1.38605(12) & 0.044282(18) \\
10   & 0.2217418(16)    & 1.55678(11) & 2.88449(46) & 1.388754(97) & 0.033232(12) \\
8    & 0.2218008(22)    & 1.546454(96) & 2.83564(39) & 1.389319(83) & 0.0233987(75) \\
6    & 0.2219509(30)    & 1.529671(87) & 2.75619(35) & 1.383783(71) & 0.0148953(38) \\
\hline
\end{tabular}
\end{center}
\end{table}

\begin{table}
\caption{The pseudocritical couplings
$\widetilde{\beta}_c$ and the values of $\chi/L^2$ (where
$\chi$ is the susceptibility) and $\partial Q/\partial \beta$ (where $Q=1/U$)
at $\beta =\widetilde{\beta}_c$
depending on the linear system size $L$.
\label{tab2} }
\begin{center}
\begin{tabular}{|c|c|c|c|}
\hline
\rule[-2mm]{0mm}{7mm}
L & $\widetilde{\beta}_c$ & $\chi/L^2$ & $10^{-3}\partial Q/\partial \beta$ \\
\hline
3456 & 0.2216546238(37)  & 1.1559(27)  & 346.2(2.9) \\
3072 & 0.2216546245(31)  & 1.1623(20) & 285.4(1.8) \\
2560 & 0.2216546231(25)  & 1.1694(12) & 216.15(72) \\
2048 & 0.2216546290(35)  & 1.1780(14) & 151.64(60) \\
1728 & 0.2216546366(43)  & 1.1889(11) & 116.53(41) \\
1536 & 0.2216546300(52)  & 1.1910(13) & 96.41(40) \\
1280 & 0.2216546329(60)  & 1.1984(11) & 71.75(21) \\
1024 & 0.221654618(12)   & 1.2068(13) & 50.27(20) \\
864  & 0.221654629(14)   & 1.2153(13) & 38.47(15)\\
768  & 0.2216546466(99)  & 1.22161(82) & 32.091(73) \\
640  & 0.221654652(19)   & 1.2287(11) & 24.034(77) \\
512  & 0.221654622(20)   & 1.23544(85) & 16.725(39) \\
432  & 0.221654640(27)   & 1.24398(83) & 12.857(27) \\
384  & 0.221654615(30)   & 1.24874(88) & 10.630(24) \\
320  & 0.221654785(35)   & 1.25899(78) & 8.015(15) \\
256  & 0.221654667(55)   & 1.26546(80) & 5.578(11) \\
216  & 0.221654511(69)   & 1.27170(73) & 4.2669(77) \\
192  & 0.221654551(85)   & 1.27619(77) & 3.5452(69) \\
160  & 0.22165455(10)    & 1.28349(72) & 2.6574(46) \\
128  & 0.22165425(14)    & 1.29143(72) & 1.8635(32) \\
108  & 0.22165391(16)    & 1.29818(63) & 1.4202(22) \\
96   & 0.22165352(19)    & 1.30099(62) & 1.1782(16) \\
80   & 0.22165313(21)    & 1.30763(51) & 0.8838(11) \\
64   & 0.22165145(29)    & 1.31468(49) & 0.61968(61) \\
54   & 0.22164984(30)    & 1.31932(40) & 0.47353(47) \\
48   & 0.22164760(37)    & 1.32176(39) & 0.39276(36) \\
40   & 0.22164346(46)    & 1.32576(38) & 0.29433(25) \\
32   & 0.22163517(56)    & 1.32880(34) & 0.20624(15) \\
27   & 0.22162278(70)    & 1.32938(34) & 0.15723(11) \\
24   & 0.22161151(71)    & 1.32989(29) & 0.130560(80)\\
20   & 0.2215823(10)     & 1.32849(28) & 0.097652(50) \\
16   & 0.2215239(12)     & 1.32483(22) & 0.068484(30) \\
12   & 0.2213817(25)     & 1.31398(29) & 0.043324(25) \\
10   & 0.2212109(32)     & 1.30356(28) & 0.032393(18) \\
8    & 0.2208676(43)     & 1.28584(23) & 0.022727(10) \\
6    & 0.2200070(59)     & 1.25323(21) & 0.0143860(60)  \\
\hline
\end{tabular}
\end{center}
\end{table}

\section{The estimation of $\omega$ in the $\xi_{\mathrm{2nd}}/L=0.6431$
scaling regime}
\label{sec:secondc}

As it is well known~\cite{HasRev,KMR17}, the dimensionless quantities related to the magnetization
cumulants as, e.~g.,  $U=\langle m^4 \rangle / \langle m^2 \rangle^2$,
$U_6=\langle m^6 \rangle / \langle m^2 \rangle^3$, etc., scale as
$A + B L^{-\omega} + \mathcal{O}\left(L^{-\bar{\omega}} \right)$
at $L \to \infty$ at the critical point $\beta=\beta_c$, as well as at
$\beta=\bar{\beta}_c$. Here $\omega$ is the leading
correction-to-scaling exponent, whereas the exponent $\bar{\omega}$ describes
the dominant subleading correction term. Note that $\bar{\beta}_c$ is
defined by the condition $\xi_{\mathrm{2nd}}/L=0.6431$, but the precise value
of $\xi_{\mathrm{2nd}}/L$ is not crucial, i.~e., it does not affect the scaling form.

\begin{figure}
\begin{center}
\includegraphics[width=0.485\textwidth]{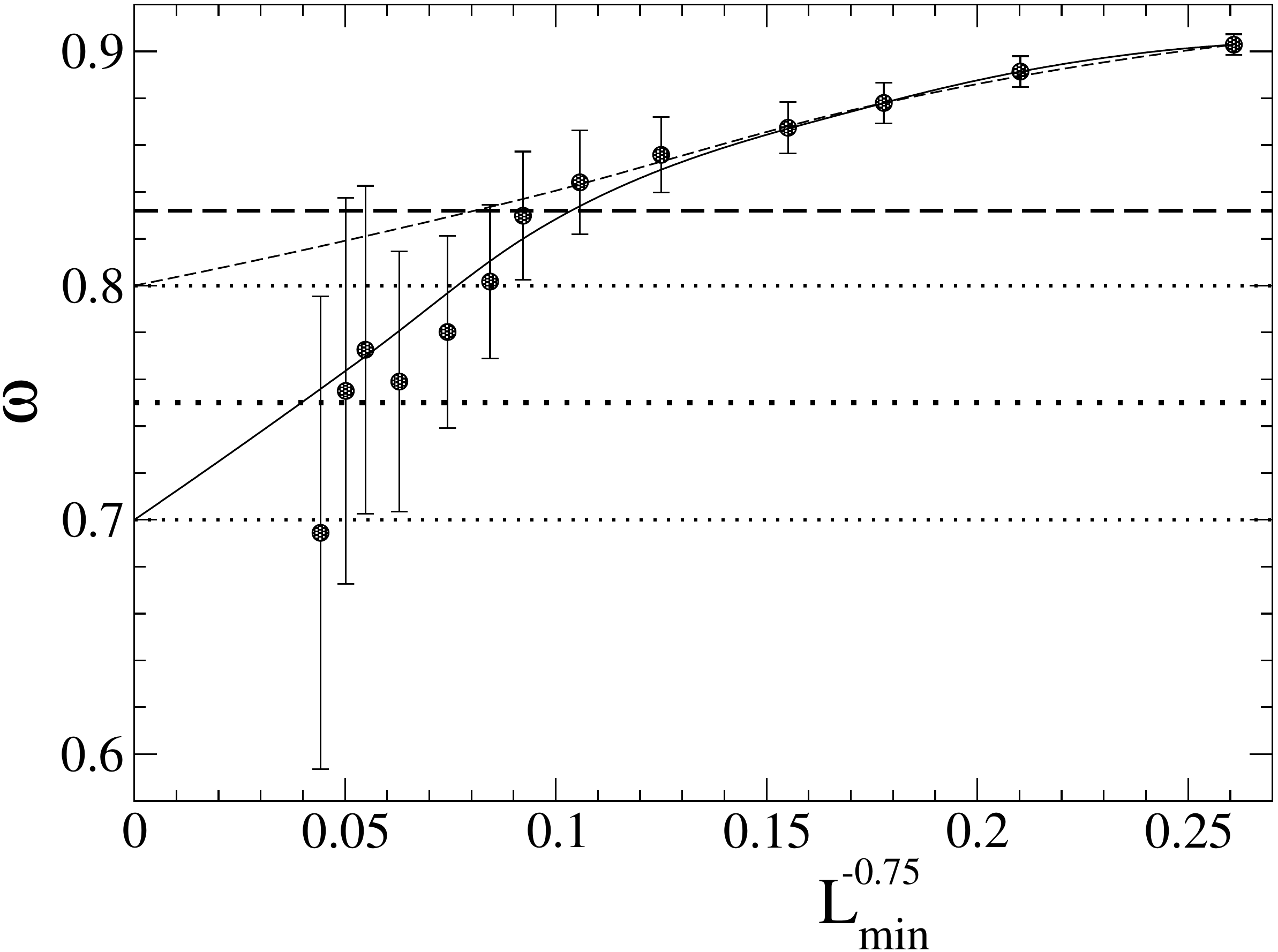}
\hfill
\includegraphics[width=0.485\textwidth]{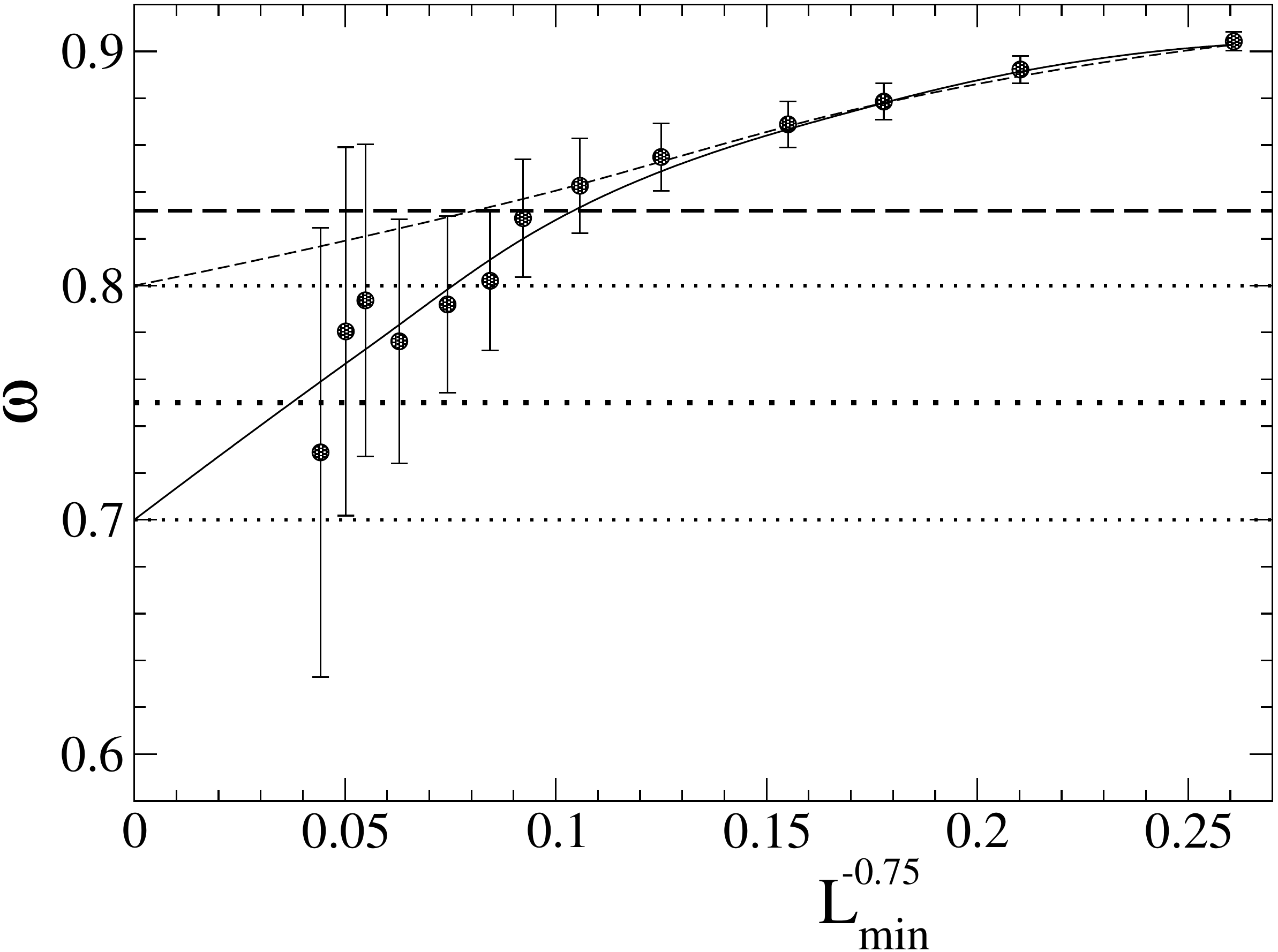}
\end{center}
\caption{The correction-to-scaling exponent $\omega$, estimated from the fit of $U$ (left) and $U_6$ (right) data to the ansatz $A + B L^{-\omega}$ within
$L \in [L_{\mathrm{min}},3072]$ and plotted as a function of $L_{\mathrm{min}}^{-0.75}$.
The solid and the dashed curves indicate the expected behavior if the asymptotic
$\omega$ value is $0.7$ and $0.8$, respectively. These two values correspond to
the lower and the upper bound of the MCRG estimate $\omega=0.75(5)$ of~\cite{Ron}, noting that $\omega \approx 0.7$ is also the MCRG estimate in~\cite{Gupta}.
The central values with the error bounds of $\omega=0.75(5)$ are shown by horizontal
dotted lines, whereas the horizontal dashed line indicates
the MC value $\omega=0.832(6)$ of~\cite{Has1}.}
\label{fig:omeff}
\end{figure}

We have estimated $\omega$ by fitting the $U$ data in Tab.~\ref{tab1} to the
ansatz $U(L)=A + B L^{-\omega}$ within $L \in [L_{\mathrm{min}},L_{\mathrm{max}}]$,
where $L_{\mathrm{min}} \in [6,64]$ and $L_{\mathrm{max}}=3072$.
The results together with those ones extracted from the $U_6(L)=A + B L^{-\omega}$ fits are shown in Fig.~\ref{fig:omeff}. The seen here $L_{\mathrm{min}}$ dependence
of the estimated
$\omega$ values is governed by the subleading correction term
$\mathcal{O}\left(L^{-\bar{\omega}} \right)$. Since $L_{\mathrm{min}} \ll L_{\mathrm{max}}$ holds, we assume that the deviations
from the correct asymptotic $\omega$ value are caused mainly by the
finiteness of $L_{\mathrm{min}}$ and not by that of $L_{\mathrm{max}}$.

We observe that our estimated exponent $\omega$ tends to decrease below the
commonly accepted MC value $0.832(6)$ of~\cite{Has1} with increasing of
$L_{\mathrm{min}}$. Therefore, we have found it interesting to test the consistency
of this behavior with the already mentioned in Sec.~\ref{intro} MCRG estimates
$\omega \approx 0.7$ and $\omega=0.75(5)$~\cite{Gupta,Ron}.
For this purpose, we have plotted the estimated $\omega$ as a function
of $L_{\mathrm{min}}^{-0.75}$, since this plot is expected to be an almost linear
function of
$L_{\mathrm{min}}^{\omega-\bar{\omega}} \sim L_{\mathrm{min}}^{-0.75}$
at large $L_{\mathrm{min}}$, because the subleading correction term
$\mathcal{O}\left(L^{-\bar{\omega}} \right)$ is by a factor
$L^{\omega-\bar{\omega}}$ smaller
than $L^{-\omega}$. The used here relation $\omega-\bar{\omega} \approx -0.75$
is true if $\bar{\omega}=2\omega$
and $\omega \approx 0.75$ hold. Here we have considered $\bar{\omega}=2 \omega$
as a plausible possibility, representing the case when the subleading
correction is, in fact, the second-order correction of the type
$\left(L^{-\omega} \right)^2=L^{-2 \omega}$. Another possibility is
$\bar{\omega}=\omega_2$, if $\omega_2<2 \omega$, where $\omega_2$ is the second irrelevant RG exponent. It is $\omega_2 = 1.67(11)$ according to~\cite{uuu}.
Thus, we generally have $\bar{\omega} = \min \{ 2 \omega,\omega_2 \}$ and
$\bar{\omega} -\omega \approx 0.75$ for $\omega \approx 0.75$.
Hence, it is meaningful to plot
the estimated $\omega$ as a  function of $L_{\mathrm{min}}^{-0.75}$
for our testing purposes.

We have drawn in Fig.~\ref{fig:omeff} spline curves, which converge asymptotically
almost linearly to certain values, i.~e., $\omega=0.7$ (solid curves) and
$\omega=0.8$ (dashed curves). The solid curves fit the data very well,
thus illustrating a plausible scenario of convergence towards
$\omega \approx 0.7$ in a good agreement with the MCRG estimation of~\cite{Gupta}.
It corresponds also to the lower bound of the MCRG estimate $\omega=0.75(5)$
of~\cite{Ron}. The upper bound of this estimate corresponds to the dashed
curves in Fig.~\ref{fig:omeff}, which fit the data marginally well.
Hence, from a purely statistical point of view, $\omega \approx 0.8$ is also
plausible and even $\omega \approx 0.83$ is possible,
but, with a larger probability, $\omega<0.8$ holds.

Note that the MC estimation in~\cite{Has1} is also based on the $U$ and
$U_6$ data at $\beta = \bar{\beta}_c$, just as here. The value
$\omega=0.832(6)$ of~\cite{Has1} has a relatively small statistical error, however,
their estimation is based on $L_{\mathrm{min}} \le 20$ and
$L_{\mathrm{max}} = 360$, whereas we have
$L_{\mathrm{min}} \le 64$ and $L_{\mathrm{max}} = 3456$.
Moreover, the range $L_{\mathrm{min}} \le 20$ corresponds to just 6
smallest--$L_{\mathrm{min}}$
data points in Fig.~\ref{fig:omeff}, from which
the deviation of $\omega$ below $0.83$ is still not seen.
We do not rule out a possibility that this deviation is
caused by statistical errors in the data.
On the other hand, such a deviation is more convincingly
confirmed by the analysis of the susceptibility data in
Sec.~\ref{sec:U}, and the agreement with the MCRG estimations,
perhaps, is also not accidental.

\section{The estimation of $\omega$ in the $U=1.6$
scaling regime}
\label{sec:U}

Here we consider the scaling regime $\beta = \widetilde{\beta}(L)$,
corresponding to $U=\langle m^4 \rangle / \langle m^2 \rangle^2 =1.6$. The precise value of $U$ is not crucial,
and one can choose $1 < U < 3$ for the scaling analysis.
Using $U=1.6$ as a near-to-critical value,
the correction-to-scaling exponent $\omega$ has been estimated
in~\cite{KMR17} from the ratios of the susceptibility $\chi(2L)/\chi(L/2)$, which scale as $A+BL^{-\omega}$ at $L \to \infty$. The result, obtained from the $\chi(L)$ data
within $L \in [40,2560]$ was $\omega=0.21(29)$. A similar estimation from the
refined data in Tab.~\ref{tab2} within $L \in [40,3456]$ gives $\omega=0.67(15)$.
Hence, one can conclude that the unusually small value of $\omega$ obtained
earlier in~\cite{KMR17} is partly due to the large statistical errors.
Therefore, the conjecture $\omega \le \omega_{\mathrm{max}}$ with
$\omega_{\mathrm{max}} \approx 0.38$, proposed earlier in~\cite{KMR17},
probably could not be confirmed. A problem here is that this conjecture
is based on a theorem proven in~\cite{our2D}, the conditions of which have been numerically verified only in the two-dimensional $\phi^4$ model.

\begin{figure}
\begin{center}
\includegraphics[width=0.75\textwidth]{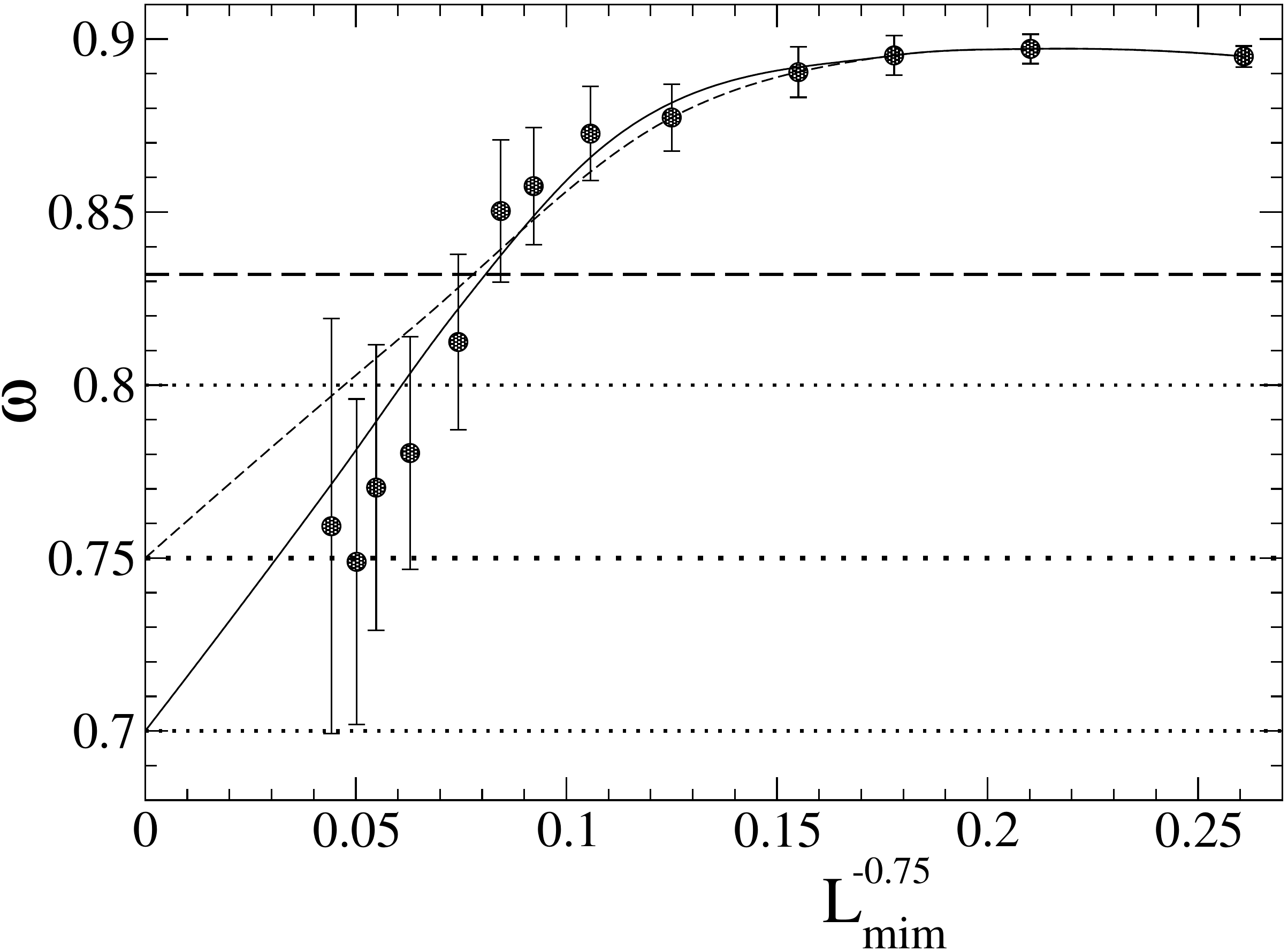}
\end{center}
\caption{The correction-to-scaling exponent $\omega$, estimated from the
fit of the susceptibility data to the ansatz~(\ref{eq:chi}) with fixed
$\eta=0.0362978(20)$ (taken from~\cite{bootstrap}) within
$L \in [L_{\mathrm{min}},3456]$ and plotted as a function of
$L_{\mathrm{min}}^{-0.75}$.
The dashed and the solid curves indicate the expected behavior if the asymptotic
$\omega$ value is $0.75$ and $0.7$, respectively. These two values correspond to
the central value and the lower bound of the MCRG estimate $\omega=0.75(5)$ of~\cite{Ron}, noting that $\omega \approx 0.7$ is also the MCRG estimate in~\cite{Gupta}.
The central values with the error bounds of $\omega=0.75(5)$ are shown by horizontal
dotted lines, whereas the horizontal dashed line indicates
the MC value $\omega=0.832(6)$ of~\cite{Has1}.}
\label{fig:omeff2}
\end{figure}

For a more accurate estimation, we have evaluated $\omega$ by fitting the susceptibility data to the ansatz
\begin{equation}
 \chi(L) = L^{2-\eta} \left( a_0 + a_1 L^{-\omega} \right)
\label{eq:chi}
 \end{equation}
with fixed critical exponent $\eta=0.0362978(20)$, provided by the conformal
bootstrap method~\cite{bootstrap}. It is well justified, since our actual estimation
of $\eta$ in Sec.~\ref{sec:etanu} perfectly agrees with this $\eta$ value.
We have fitted the data within $L \in [L_{\mathrm{min}},3456]$ and have plotted
in Fig.~\ref{fig:omeff2} the resulting $\omega$ estimates as a function of
$L_{\mathrm{min}}^{-0.75}$ in order to
compare them with the MCRG values, based on the same arguments as in Sec.~\ref{sec:secondc}.

According to the solid curve in Fig.~\ref{fig:omeff2},
the behavior of $\omega$ well agrees
with the $\omega \approx 0.7$ estimate in~\cite{Gupta}, which is also
the lower bound of the $\omega=0.75(5)$ estimate in~\cite{Ron}. This behavior
is less consistent with the central value $0.75$ of the latter estimation --
see the dashed line in Fig.~\ref{fig:omeff2}, suggesting  that this
central value could be slightly overestimated rather than underestimated.

Up to now, we have only tested the consistency of $\omega$ vs
$L_{\mathrm{min}}$ behavior with the values of MCRG. For an independent estimation,
we have fit the $\chi(L)$ data to a refined ansatz
\begin{equation}
 \chi(L) = L^{2-\eta} \left( a_0 + a_1 L^{-\omega}
 + a_2 L^{-\bar{\omega}} \right) \,,
\label{eq:chi2}
\end{equation}
where $\eta=0.0362978(20)$, as before, and $\bar{\omega}$ is also fixed.
The meaning of $\bar{\omega} = \min \{ 2 \omega,\omega_2 \}$ has been already
discussed in Sec.~\ref{sec:secondc}. The main advantage of~(\ref{eq:chi2}) in
comparison with~(\ref{eq:chi}) is that it gives more stable values of
$\omega$, which do not essentially change with $L_{\mathrm{min}}$ for
$L_{\mathrm{min}} \ge 20$. Moreover, the inclusion of the
correction term $a_2 L^{-\bar{\omega}}$ improves the quality of the fit,
reducing the value of $\chi^2/\mathrm{d.o.f.}$ of the fit from $1.46$
to $1.24$ at $\bar{\omega}=1.5$ and $L_{\mathrm{min}} = 20$.
We have set $L_{\mathrm{min}} = 20$ as the best choice, since it gives
the smallest statistical error among the fits with $L_{\mathrm{min}} \ge 20$.
The fitting results are collected in Tab.~\ref{tab3}.
\begin{table}
\caption{The values of $\omega$ depending on $\bar{\omega}$, obtained by fitting the susceptibility
data to the ansatz~(\ref{eq:chi2}) within $L \in [20,3456]$ at
$\eta=0.0362978(20)$. The $\chi^2/\mathrm{d.o.f.}$ of the fit
is indicated in the third column.
\label{tab3} }
\begin{center}
\begin{tabular}{|c|c|c|}
\hline
\rule[-2mm]{0mm}{7mm}
$\bar{\omega}$ & $\omega$ & $\chi^2/\mathrm{d.o.f.}$ \\
\hline
1.7 & 0.718(56)  & 1.24  \\
1.6 & 0.701(61)  & 1.24  \\
1.5 & 0.679(68)  & 1.24  \\
1.4 & 0.651(76)  & 1.24  \\
\hline
\end{tabular}
\end{center}
\end{table}
Evidently, the estimated $\omega$ decreases with decreasing of $\bar{\omega}$.
Taking into account
the statistical error bars of one $\sigma$ and choosing $\bar{\omega}$
self-consistently, i.~e.,
$\bar{\omega} \le 2 \omega$, our fitting procedure suggests
that $\omega<0.75$ most probably holds. More precisely, $\omega<0.75$
holds if our fit at the correct value of $\bar{\omega}$
does not underestimate $\omega$ by more than one standard error $\sigma$.
Recall that the possibility $\omega<0.75$ is supported also by the behavior
seen in Fig.~\ref{fig:omeff2}.

\section{A critical re-examination of MCRG data}
\label{sec:reanal}

For a more complete picture,
we have reanalysed the MCRG data of~\cite{Ron},
providing some original idea about the estimation of $\omega$
from the MCRG iterations. In~\cite{Ron},
one starts with some lattice size $L$ and makes $n$ renormalization
iterations. In each iteration, the lattice size shrinks by the factor $b=2$,
so that the final lattice size is $L_f = 2^{-n} L$. Our idea is to look
on the MCRG data in a somewhat different way than the authors
of~\cite{Ron} originally did.
Instead of considering a sequence with fixed $L$ and $n=1,2, \ldots$, we propose
to look on a sequence of data with fixed final lattice size $L_f$ and $n=1,2, \ldots$
It ensures that the finite-size effects do not increase along such
a sequence. We can fit one such sequence to evaluate the considered
critical exponent at $n \to \infty$ at a given $L_f$. Then we can compare the
fit results for sequences with different $L_f$, say, $L_f=8$ and $L_f=16$,
to make a finite-size correction on $L_f$.

\begin{figure}
\begin{center}
\includegraphics[width=0.75\textwidth]{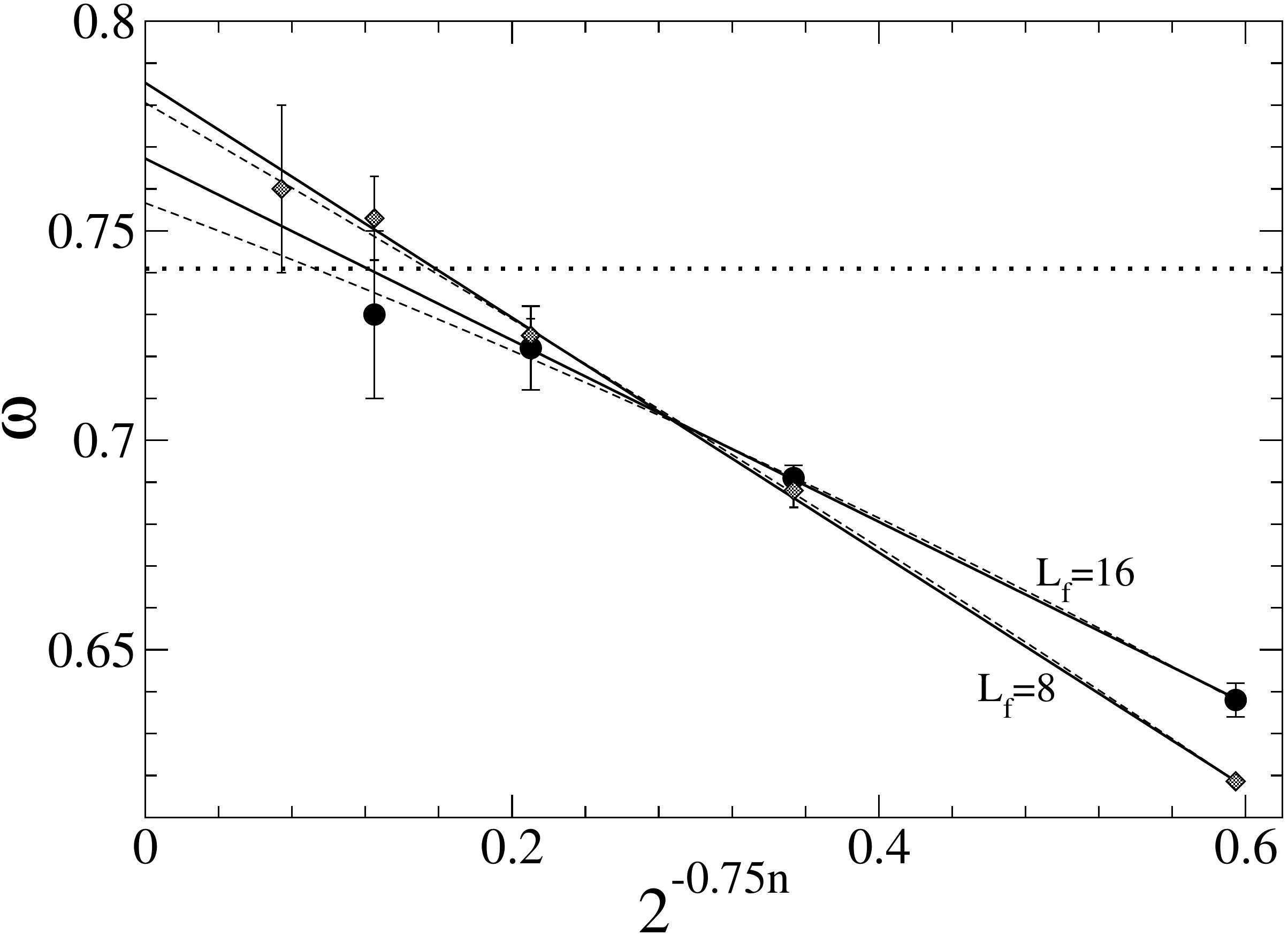}
\end{center}
\caption{The correction-to-scaling exponent $\omega$ depending on $2^{-0.75n}$
according to the  MCRG data in~\cite{Ron},
where $n$ is the number of MCRG iterations, starting with the lattice
size $L=2^n L_f$ and ending up with the lattice size $L_f$. The data with
$L_f=8$ and $L_f=16$ are shown by diamonds and circles, respectively.
The linear (solid lines) and quadratic (dashed lines) fits
show the estimated convergence to certain $\omega$ values at $n \to \infty$
for the given $L_f$. The horizontal dotted line represents the
estimated asymptotic (at $n \to \infty$) value $0.741(21)$
of $\omega$ at $L_f=\infty$.
}
\label{fig:MCRG}
\end{figure}

We have applied this method to the estimation of $\omega$, using the
$-y_{T2} \equiv \omega$ data in Tab.~V of~\cite{Ron} at the maximal
number of operators $N_e=30$ considered there.
Following the idea in~\cite{Gupta}, we have plotted $\omega$ as a
function of $2^{-\omega n}$ in Fig.~\ref{fig:MCRG}, using here $0.75$ as an approximate self-consistent value of $\omega$. It allows us
to estimate the asymptotic $\omega$ at $n \to \infty$ by a linear or a quadratic extrapolation.
Apparently, there is some finite-size effect
in this estimation, since the linearly extrapolated $\omega$ values
(see the solid lines in Fig.~\ref{fig:MCRG})
slightly depend on $L_f$. Namely, we have
$\omega=0.7853(48)$ at $L_f=8$ and $\omega=0.7672(80)$ at $L_f=16$.
The finite-size correction is expected to be
$\sim L_f^{-\omega}$ with $\omega \approx 0.75$. The corresponding linear extrapolation of $\omega(L_f)$ as a function of $L^{-0.75}$
gives the estimate $\omega=0.741(21)$ for $L_f=\infty$.\
This value is indicated in Fig.~\ref{fig:MCRG} by the horizontal dotted line.
This estimation perfectly agrees with that in~\cite{Ron}, but
the statistical error bars are significantly smaller. Thus, one could
judge that $\omega$ is, indeed, significantly smaller than $0.83$.

However, there are still uncertainties
about the possible systematic errors, as pointed out below.
\begin{enumerate}
 \item
 The evaluation in Fig.~\ref{fig:MCRG} is based on linear fits
 (the solid straight-line fits),
but the data plots may have some curvature (see the dashed fit curves in
Fig.~\ref{fig:MCRG}). Unfortunately, this curvature cannot
be reliably estimated because of too large statistical errors.
 \item
There is an uncertain systematical error caused by the inaccuracy
in the value of the critical coupling $\beta_c$, used in the simulations
by Ron et.~al.~\cite{Ron}. Indeed, the used there value $0.2216544$
is slightly underestimated.
A more precise value is $0.22165462$, as it can be seen from
Tabs.~\ref{tab1} and~\ref{tab2}, as well as from a recent MC estimation
in~\cite{Ferrenberg}.
\item
The number of operators has been limited to $N_e \le 30$ in~\cite{Ron}, not even including all those ones considered in~\cite{Gupta}.
A larger number of operators might be important.
\end{enumerate}

These uncertainties should be resolved for a reliable estimation of
the critical exponents, including $\omega$.

\section{Estimation of the critical exponents $\eta$ and $\nu$}
\label{sec:etanu}

We have estimated the critical exponent $\eta$ by fitting the susceptibility
data in Tabs.~\ref{tab1} and~\ref{tab2} to the ansatz~(\ref{eq:chi}).
The fitting results, depending on the minimal lattice size $L_{\mathrm{min}}$
included in the fit, are collected in Tabs.~\ref{tab4} and~\ref{tab5},
respectively. In these tables, $\eta_0$ is the estimate of $\eta$
without corrections to scaling, obtained by formally setting $a_1=0$ in~(\ref{eq:chi}).
The other two estimates $\eta_1$ and $\eta_2$ are obtained by fitting
the data to the full ansatz~(\ref{eq:chi}) with fixed
exponent $\omega=0.75$ (for $\eta_1$) or $\omega=0.832$ (for $\eta_2$).
The results for $\beta=\bar{\beta}(L)$ in Tab.~\ref{tab4} appear to be
more accurate than those for $\beta=\widetilde{\beta}(L)$ in Tab.~\ref{tab5}.
Besides, the results  without corrections  to scaling ($\eta_0$) show
some trend, therefore the inclusion of the correction term is meaningful.
It allows to fit the data reasonably well already starting
with $L_{\mathrm{min}}=16$. Moreover, it stabilises the $\eta_1$ and $\eta_2$
values in such a way
that the estimates at $L_{\mathrm{min}}=16$ and $L_{\mathrm{min}}>16$ perfectly
match. Since $L_{\mathrm{min}}=16$ ensures the smallest statistical error
in this case, our final result ($\eta_1$ from Tab.~\ref{tab4}) is
$\eta = 0.03632(13)$. Here we have chosen $\omega=0.75$, as it well
coincides with our estimations in Secs.~\ref{sec:secondc} and~\ref{sec:U},
but $\omega=0.832$ gives practically the same result $\eta=0.03630(12)$
($\eta_2$ from Tab.~\ref{tab4}).
Note also that the $\eta_1$ and $\eta_2$ estimates in Tab.~\ref{tab5} perfectly
agree with those in Tab.~\ref{tab4}, only the statistical error bars are larger.

\begin{table}
\caption{The MC estimates $\eta_0$, $\eta_1$ and $\eta_2$ of the critical
exponent $\eta$, obtained by fitting
the susceptibility data at $\beta=\bar{\beta}_c(L)$ (Tab.~\ref{tab1})
within $L \in [L_{\mathrm{min}},3072]$ at different values of
$L_{\mathrm{min}}$. The estimates $\eta_0$
are provided by the fits without corrections to scaling
(setting $a_1=0$ in~(\ref{eq:chi})), whereas $\eta_1$ and $\eta_2$
are obtained by fitting the data to~(\ref{eq:chi})
with fixed $\omega=0.75$ and $\omega=0.832$, respectively.
The $\chi^2/\mathrm{d.o.f.}$ of the fits for  $\eta_0$, $\eta_1$ and $\eta_2$
are given in the columns Nr.~3, 5 and 7, respectively.
\label{tab4} }
\begin{center}
\begin{tabular}{|c|c|c||c|c||c|c|}
\hline
\rule[-2mm]{0mm}{7mm}
$L_{\mathrm{min}}$ & $\eta_0$ & \!$\chi^2/\mathrm{d.o.f}$\! & $\eta_1$ &
\!$\chi^2/\mathrm{d.o.f}$\! & $\eta_2$ & \!$\chi^2/\mathrm{d.o.f}$\! \\
\hline
16 & 0.037295(66) & 3.60 & 0.03632(13) & 1.20 & 0.03630(12) & 1.20 \\
20 & 0.037062(77) & 2.43 & 0.03625(16) & 1.21 & 0.03622(14) & 1.21 \\
24 & 0.036877(85) & 1.56 & 0.03636(18) & 1.20 & 0.03631(16) & 1.20 \\
27 & 0.036764(94) & 1.29 & 0.03644(20) & 1.20 & 0.03638(17) & 1.20 \\
32 & 0.03670(11)  & 1.28 & 0.03642(22) & 1.25 & 0.03637(19) & 1.25 \\
40 & 0.03664(12)  & 1.28 & 0.03642(27) & 1.30 & 0.03636(23) & 1.31 \\
48 & 0.03653(14)  & 1.21 & 0.03669(31) & 1.25 & 0.03659(27) & 1.25 \\
\hline
\end{tabular}
\end{center}
\end{table}

\begin{table}
\caption{The MC estimates $\eta_0$, $\eta_1$ and $\eta_2$ of the critical
exponent $\eta$, obtained by fitting
the susceptibility data at $\beta=\widetilde{\beta}_c(L)$ (Tab.~\ref{tab2})
within $L \in [L_{\mathrm{min}},3456]$ at different values of
$L_{\mathrm{min}}$. The estimates $\eta_0$
are provided by the fits without corrections to scaling
(setting $a_1=0$ in~(\ref{eq:chi})), whereas
$\eta_1$ and $\eta_2$
are obtained by fitting the data to~(\ref{eq:chi})
with fixed $\omega=0.75$ and $\omega=0.832$, respectively.
The $\chi^2/\mathrm{d.o.f.}$ of the fits for  $\eta_0$, $\eta_1$ and $\eta_2$
are given in the columns Nr.~3, 5 and 7, respectively.
\label{tab5} }
\begin{center}
\begin{tabular}{|c|c|c||c|c||c|c|}
\hline
\rule[-2mm]{0mm}{7mm}
$L_{\mathrm{min}}$ & $\eta_0$ & \!$\chi^2/\mathrm{d.o.f}$\! & $\eta_1$ &
\!$\chi^2/\mathrm{d.o.f}$\! & $\eta_2$ & \!$\chi^2/\mathrm{d.o.f}$\! \\
\hline
16 & 0.03807(15) & 2.93 & 0.03633(29) & 1.32 & 0.03599(26) & 1.31 \\
20 & 0.03768(17) & 2.32 & 0.03600(35) & 1.25 & 0.03574(30) & 1.27 \\
24 & 0.03737(19) & 1.87 & 0.03595(40) & 1.30 & 0.03570(35) & 1.31 \\
27 & 0.03719(21) & 1.77 & 0.03585(43) & 1.34 & 0.03561(38) & 1.35 \\
32 & 0.03682(23) & 1.36 & 0.03615(48) & 1.31 & 0.03587(42) & 1.32 \\
40 & 0.03655(27) & 1.27 & 0.03646(58) & 1.32 & 0.03615(50) & 1.34 \\
48 & 0.03649(30) & 1.31 & 0.03665(67) & 1.37 & 0.03632(58) & 1.38 \\
\hline
\end{tabular}
\end{center}
\end{table}

We have estimated the critical exponent $\nu$ by fitting the
$\partial Q/ \partial \beta$ data (where $Q=1/U$) at $\beta=\widetilde{\beta}_c(L)$
(see Tab.~\ref{tab2})
to the ansatz
\begin{equation}
 \frac{\partial Q}{\partial \beta} = L^{1/\nu}
 \left( a_0 + a_1 L^{-\omega} \right)
 \label{eq:Qat}
\end{equation}
within $L \in [L_{\mathrm{min}},3456]$. The results are collected in Tab.~\ref{tab6}.
The $\partial Q/ \partial \beta$ data at $\beta=\bar{\beta}_c(L)$ can also be used, however, these data provide fits of significantly lower quality
(with $\chi^2/\mathrm{d.o.f.} \gtrsim 1.83$) and somewhat larger statistical
errors.

\begin{table}
\caption{The MC estimates $\nu_0$, $\nu_1$ and $\nu_2$ of the critical
exponent $\nu$, obtained by fitting
the $\partial Q/\partial \beta$ data at $\beta=\widetilde{\beta}_c(L)$ (Tab.~\ref{tab2})
within $L \in [L_{\mathrm{min}},3456]$ at different values of
$L_{\mathrm{min}}$.
The estimates $\nu_0$
are provided by the fits without corrections to scaling (setting
$a_1=0$ in~(\ref{eq:Qat})), whereas
$\nu_1$ and $\nu_2$
are obtained by fitting the data to~(\ref{eq:Qat})
with fixed $\omega=0.75$ and $\omega=0.832$, respectively.
The $\chi^2/\mathrm{d.o.f.}$ of the fits for  $\nu_0$, $\nu_1$ and $\nu_2$
are given in the columns Nr.~3, 5 and 7, respectively.
\label{tab6} }
\begin{center}
\begin{tabular}{|c|c|c||c|c||c|c|}
\hline
\rule[-2mm]{0mm}{7mm}
$L_{\mathrm{min}}$ & $\nu_0$ & \!$\chi^2/\mathrm{d.o.f}$\! & $\nu_1$ &
\!$\chi^2/\mathrm{d.o.f}$\! & $\nu_2$ & \!$\chi^2/\mathrm{d.o.f}$\! \\
\hline
16 & 0.63048(19) & 1.35 & 0.63026(38) & 1.38 & 0.63028(33) & 1.38 \\
20 & 0.63047(22) & 1.40 & 0.63008(45) & 1.41 & 0.63011(39) & 1.41 \\
24 & 0.63041(25) & 1.44 & 0.62999(52) & 1.46 & 0.63002(45) & 1.46 \\
27 & 0.63046(27) & 1.49 & 0.62969(56) & 1.45 & 0.62975(49) & 1.44 \\
32 & 0.63017(31) & 1.38 & 0.63008(63) & 1.44 & 0.63005(55) & 1.43 \\
40 & 0.62996(36) & 1.38 & 0.63075(75) & 1.38 & 0.63061(65) & 1.38 \\
48 & 0.63011(40) & 1.42 & 0.63080(88) & 1.44 & 0.63067(76) & 1.45 \\
\hline
\end{tabular}
\end{center}
\end{table}

The $\nu$ estimates, denoted by $\nu_0$ in Tab.~\ref{tab6},
are obtained neglecting the
correction term $a_1 L^{-\omega}$ in~(\ref{eq:Qat}), whereas
$\nu_1$ and $\nu_2$ are the estimates with fixed $\omega=0.75$ and
$\omega=0.832$, respectively. In fact, we can see from Tab.~\ref{tab6}
that the inclusion of the correction term
neither improves the quality of the fits nor remarkably changes
the fitting results. Hence, the estimation
without corrections to scaling is appropriate in this case.
The estimate $\nu =\nu_0= 0.63048(19)$ at $L_{\mathrm{min}}=16$
has the smallest statistical error and the corresponding fit has the smallest
$\chi^2/\mathrm{d.o.f.}$ value. On the other hand, this $\nu_0$ value
is the largest one among those listed in Tab.~\ref{tab6}.
Therefore, to avoid a possible tiny overestimation because of a too small value of $L_{\mathrm{min}}$,
we have assumed $\nu=0.63017(31)$, obtained at $L_{\mathrm{min}}=32$,
as our final estimate. This value closely agrees with all other estimates in Tab.~\ref{tab6}.

Summarising our MC estimation of the critical exponents $\eta$ and $\nu$,
we note that our final estimates  $\eta=0.03632(13)$ and $\nu=0.63017(31)$
perfectly agree with the ``exact'' values of the conformal bootstrap
method, i.~e., $\eta=0.0362978(20)$ and $\nu=0.6299709(40)$~\cite{bootstrap}.

\section{Discussion and outlook}
\label{sec:discussion}

The critical exponents
$\eta=0.03632(13)$ and $\nu=0.63017(31)$
of our MC analysis of very large lattices ($L \le 3456$)
agree well with the known CFT values~\cite{bootstrap},
which are claimed to be extremely accurate or ``exact''.
It rises the confidence about correctness of our MC simulations
and analysis.

Our MC analysis suggests that the correction-to-scaling
exponent $\omega$
could be smaller than the commonly accepted value about $0.83$,
thus supporting
the MCRG estimates $\omega \approx 0.7$~\cite{Gupta}
and $\omega=0.75(5)$~\cite{Ron}.
Of course, due to the statistical errors, our fits can appear to be wrong
by more than one $\sigma$. In fact, the fit to~(\ref{eq:chi2}) should be wrong by
$\approx 2 \sigma$ to reach the consistency with $\omega=0.832(6)$ in~\cite{Has1}
or $\omega=0.82951(61)$ in~\cite{CFTrecent}.
This is possible, although the agreement with the MCRG estimates
allows us to think that the observed deviation of $\omega$ below $0.83$
with increasing of $L_{\mathrm{min}}$ { in Figs.~\ref{fig:omeff} and~\ref{fig:omeff2}} is a real effect.
There is still a question whether or not this deviation represents the true asymptotic behavior. Indeed,
we cannot exclude a possibility that the estimated $\omega$ values would
increase, being accurately extracted from the data for even larger lattice sizes.
To test this scenario, accurate enough data for very large lattice sizes
should be obtained. It requires an enormous computational effort.
{ Theoretically, the asymptotic $\omega$ value could
appear to be larger due to some correction term(s), which are not yet
included in~(\ref{eq:chi2}). For example,
$\propto L^{-1}$ correction
term in the brackets of~(\ref{eq:chi2})
could significantly influence
the results because the exponent $1$ is close to $0.83$.
Unfortunately, we currently do not have any theoretical argument for
the existence of such a correction term in the 3D Ising model.
One could just try to make fits with extra terms included in~(\ref{eq:chi2}).
From a technical point of view, however, such fits
are problematic, as they require higher accuracy of the data.

Taking into account the computational efforts required for the above discussed
improved MC analysis, a refining of MCRG estimation, in order to control and eliminate
the systematical errors discussed in Sec.~\ref{sec:reanal},
is a more feasible task.}
If $\omega$ is, indeed, about $0.83$, then we should see the
convergence of the MCRG iterations to this value.
Based on a systematic and rigorous analysis of the MCRG data
(as, e.~g., proposed in Sec.~\ref{sec:reanal}), sufficiently
accurate and reliable estimates of $\omega$ could be easily obtained,
slightly extending the simulations, e.~g., to $L=512$, if necessary.

One could judge that $\omega=0.82951(61)$ holds according
to the CFT~\cite{Showk,CFTrecent}, therefore
the considered here significantly smaller $\omega$ values
describe, in the best case, a transient behavior.
On the other hand, the possibility of $\omega<0.8$ cannot be rigorously
excluded because of the following issues.
\begin{itemize}
 \item
 The CFT is an asymptotic theory.
 Hence, it should correctly describe the leading scaling behavior,
 but not necessarily corrections to scaling.
 { Indeed, the MC analysis in~\cite{our2D} shows the
 existence of non-integer correction-to-scaling exponents
 in the scalar 2D $\phi^4$ model, which are not expected from the CFT.
 This idea is supported also by the resummed $\varepsilon$-expansion
 in~\cite{GZJ}. It yields $\omega=1.6 \pm 0.2$ for this
 model, simultaneously providing accurate results for the known
 critical exponents $\gamma$, $\nu$, $\beta$ and $\eta$ in two dimensions.
 In the 3D Ising model, the subleading correction-to-scaling exponent
 $\omega_2=1.67(11)$, extracted from the nonperturbative RG analysis
 in~\cite{uuu}, apparently does not coincide with the
 results of the CFT in~\cite{CFTrecent}. Namely,
 we find $\omega_2=3.8956(43)$ in accordance with
 $\Delta_{\epsilon''} = 6.8956(43)$ in~\cite{CFTrecent} and the relation
$\Delta_{\epsilon''} = 3 + \omega_2$ given in~\cite{Showk}.
This might be a serious issue,} although it still does not rule out a possibility
that the CFT could correctly capture the leading corrections
to scaling in the 3D Ising model.
One could mention that the estimate
$\Delta_{\epsilon''} = 6.8956(43)$ is
not rigorous, but nevertheless is considered as ``somewhat accurate'' in~\cite{CFTrecent}.

 \item Only a subset of all known estimates of $\omega$ in the 3D Ising
 model well coincide with $\omega=0.82951(61)$. The purely numerical estimations
 are usually considered as being most reliable. The MC estimates here
 and in~\cite{Has1} do not well agree. The high temperature series expansion in~\cite{HTCompostrini} gives $\omega=\Delta/\nu=0.825(48)$
 and confirms $\omega=0.82951(61)$ within the error bars.
 On the other hand, a more recent
 estimation by the large mass expansion in~\cite{Yamada} gives
 $\omega \approx 0.8002$. The error bars are not stated here, but
 one can judge from the series of estimates of the order $N \in [21,25]$,
 i.~e., 0.92800, 0.86046, 0.82024, 0.80411, 0.80023, that
 $\omega$, probably, is about $0.80$ and significantly smaller than $0.83$.
 Among other results, the perturbative RG estimates
 $\omega=0.782(5)$~\cite{PS08} (from the expansion at fixed dimension $d=3$)
 and $\omega=0.82311(50)$~\cite{Shalaby} (from the $\varepsilon$-expansion)
 can be mentioned, as they are claimed to be very accurate.
 The first one could be less accurate due to the
 singularity of the Callan-Symanzik $\beta$-function~\cite{PVx}.
 The estimates from the truncated nonperturbative RG equations
 are $\omega=0.870(55)$ and $\omega=0.832(14)$, obtained in~\cite{WshPRE}
 from the derivative expansion at the $O \left(\partial^{2m} \right)$
 order with $m=1$ and $m=2$, respectively. The stated here
 error bars, however, include only the uncertainty of estimation
 at the given order. Thus, the expected final result
 at $m \to \infty$ is still unclear.
 \end{itemize}

{ In summary, the MC and MCRG data analysed in this paper suggest
that the correction-to-scaling exponent $\omega$ of the 3D Ising model
could be somewhat smaller than usually expected. However, we do
not have enough data to be sure.
The precise value of $\omega$
is a quantity which merits a further consideration and testing.}
In particular, it would be very meaningful to resolve the problems
and challenges of the MCRG simulations outlined in this paper.
As we believe, it would lead to sufficiently accurate and reliable
estimates of $\omega$. In view of the above discussion, it would be
also very interesting to evaluate the subleading
correction-to-scaling exponent $\omega_2$ from the MCRG simulations.

\section*{Acknowledgments}

This work was made possible by the facilities of the
Shared Hierarchical Academic Research Computing Network
(SHARCNET:www.sharcnet.ca). 
The authors acknowledge the use of resources provided by the
Latvian Grid Infrastructure  and High Performance Computing centre
of Riga Technical University.
R. M. acknowledges the support from the
NSERC and CRC program.
{ We thank Prof. J. H. H. Perk for a discussion.}

\end{document}